# Superconducting proximity effect in a strongly correlated charge-transfer insulator

Mengya Ren[1,2,†], Yaoyao Chen[1,2,†], Fudi Zhou[1,2], Can Zhang[1,2], Zhaoteng Dong[1,2], Lili Zhou[1,2], Quanzhen Zhang[1], Huixia Yang[1], Xiaolong Xu[1], Yuanxiao Ma[1], Yu Zhang[1,2,*], Yeliang Wang[1,*]

[1]School of Integrated Circuits and Electronics, MIIT Key Laboratory for Low-Dimensional Quantum Structure and Devices, Beijing Institute of Technology, Beijing 100081, China.

[2]School of Interdisciplinary Science, State Key Laboratory of Environment Characteristics and Effects for Near-space, Beijing Key Laboratory of Quantum Matter State Control and Ultra-Precision Measurement Technology, Beijing Institute of Technology, Beijing 100081, China.

[†]These authors contributed equally to this work.

[*]Corresponding author: Yu Zhang (yzhang@bit.edu.cn); Yeliang Wang (yeliang.wang@bit.edu.cn).

**Proximity-induced superconductivity in strongly correlated insulators provides a versatile route for engineering quantum states of matter and artificial systems with tailored functionalities. However, microscopic interplay between superconductivity and correlated insulating states remains poorly understood. Here we use ultralow-temperature scanning tunnelling microscopy (STM) to systemically investigate superconducting proximity effects in a charge-transfer insulator. Via STM tip manipulation, atomically sharp lateral junctions composed of superconducting monolayer H-$NbSe_2$ and charge-transfer insulating monolayer T-$NbSe_2$ are constructed, enabling direct access to tunable coupling regimes. In the weak-coupling regime, there is a robust proximity-induced superconducting gap in T-$NbSe_2$, with a reduced gap value relative to that of H-$NbSe_2$. Upon entering the strong-coupling regime, T-$NbSe_2$ exhibits a superconducting gap comparable to that of H-$NbSe_2$, accompanied by pronounced particle-hole-symmetric in-gap bound states, consistent with Yu-Shiba-Rusinov-like excitations. These findings establish monolayer H/T-$NbSe_2$ lateral junctions as a model platform for elucidating superconducting proximity effects in strongly correlated charge-transfer insulators.**

Superconducting proximity effect is one of the central phenomena in hybrid junctions, which enables superconducting correlations to be induced in intrinsically non-superconducting materials. This effect provides a versatile route for engineering quantum states of matter and artificial systems with tailored quantum functionalities. In conventional metal/superconductor junctions, proximity-induced superconductivity originates from Andreev reflection processes at the interface, through which Cooper-pairs penetrate the normal metal over characteristic length scales that can reach several tens to hundreds of nanometres [1-5]. When the proximitized material is magnetic, superconductivity becomes intertwined with exchange fields and spin-dependent scattering, allowing spin-singlet Cooper pairs to acquire spin-triplet components, generating Zeeman-split quasiparticle spectra, and potentially stabilizing topological superconductivity [6-9].

Extending this paradigm to strongly correlated electronic systems opens an even richer frontier, where superconductivity is expected to intertwine with charge localization, magnetism, and electronic correlations to generate emergent phenomena [10,11]. However, direct experimental access to this regime remains challenging, since it requires atomically clean and electronically well-defined interfaces, together with nanoscale probes capable of resolving the electronic characteristics. As a result, microscopic understanding of superconducting proximity effects in strongly correlated electronic systems remains limited.

Lateral junctions composed of monolayer transition-metal dichalcogenides (TMDs) provide an appealing platform for elucidating superconducting proximity effect in strongly correlated electronic systems. Their layered atomic structure enables the fabrication of atomically sharp interfaces [12-14], while their rich polymorphism gives rise to diverse many-body ground states within closely related crystal frameworks. $NbSe_2$ is a prototypical example in this regard. Monolayer H-$NbSe_2$ is a charge-density-wave (CDW) metal that hosts Ising superconductivity at the tempearature below Tc ~ 3 K [15,16], whereas monolayer T-$NbSe_2$ is a strongly correlated charge-transfer insulator exhibiting both CDW order and quantum spin liquid state [17-19]. The near lattice matching between the two phases makes monolayer H/T-$NbSe_2$ lateral junctions an ideal platform for investigating the superconducting proximity effects in strongly correlated electronic systems.

In this Letter, we use ultralow-temperature scanning tunnelling microscopy (STM) to atomically probe the superconducting proximity effect in a strongly correlated charge-transfer insulator. Via the STM tip manipulation, we assemble atomically sharp monolayer H/T-$NbSe_2$ lateral junctions that provide direct access to this regime. High-resolution spectroscopic measurements reveal that the induced superconductivity in T-$NbSe_2$ is controlled by the exchange coupling between local magnetic moments in T-$NbSe_2$ and the superconductivity in H-$NbSe_2$. For weak coupling, T-$NbSe_2$ develops a robust and markedly reduced superconducting gap. In contrast, for strong coupling, particle-hole-symmetric bound states emerge within the induced superconducting gap, consistent with Yu-Shiba-Rusinov (YSR)-like excitations.

In our experiments, high-quality monolayer $NbSe_2$ islands with coexisting H and T phases were directly synthesized on bilayer-graphene-covered SiC(0001) substrates via a molecular beam epitaxy (MBE) method, as described in Methods and Fig. S1 of the Supplemental Material [20]. H-$NbSe_2$ and T-$NbSe_2$ adopt distinct trigonal-prismatic and octahedral coordination geometries, respectively, as illustrated in Figs. 1(a) and (b). In both phases, the topmost Se atoms dominate the STM topographic contrast [17,21], revealing unambiguously distinguishable CDW characteristics [15,17-19]: monolayer H-$NbSe_2$ hosts a $3 \times 3$ CDW superlattice aligned with the atomic lattice, whereas monolayer T-$NbSe_2$ hosts a $\sqrt{13} \times \sqrt{13}\ R13.9°$ CDW superlattice relative to the atomic lattice, with each supercell featuring a star-of-David (SOD) motif, as shown in Figs. 1(c), (d), and Fig. S2 [20].

Figures 1(e) and (f) present the measured electronic properties of monolayer $NbSe_2$ by means of ultralow-temperature scanning tunnelling spectroscopy (STS) at 350 mK. Monolayer H-$NbSe_2$ exhibits metallic behavior, accompanied by a striking dip near the Fermi level $E_F$, suggesting a partial gap associated with the CDW state. A closer inspection further reveals the opening of a superconducting gap symmetric about $E_F$, with a magnitude of approximately 1.8 meV (Fig. 1(e)). In stark contrast, monolayer T-

$NbSe_2$ displays charge-transfer insulating behavior, characterized by an upper Hubbard band (UHB) located at about 0.2 eV, and a lower Hubbard band (LHB) hybridized with the valence band (VB) at about -0.2 eV (Fig. 1(f)). These observations are in good agreement with previous studies [15,17,22,23].

To access the superconducting proximity effect in a strongly correlated charge-transfer insulator, we controllably construct atomically sharp monolayer H/T-$NbSe_2$ lateral heterophase homojunctions *in situ* using STM tip manipulation. Two types of junctions with distinct interfacial coupling strengths are realized within the same material system. The first is formed by tip-pushed lateral displacement of isolated H-$NbSe_2$ and T-$NbSe_2$ islands into close proximity (type-1). In this configuration, the atomic lattice is discontinuous across the interface, leading to weak electronic coupling of the superconducting H-$NbSe_2$ and the charge-transfer-insulating T-$NbSe_2$. The second is created through a tip-pulse-induced local phase transition from H-$NbSe_2$ to T-$NbSe_2$, which yields an atomically continuous H/T-$NbSe_2$ junction and thereby enables strong electronic coupling (type-2). This controllable platform enables a direct comparison between weakly and strongly coupled superconductor/charge-transfer-insulator interfaces.

Figures 2(a) and (b) schematically illustrate the manipulation procedure used to construct the type-1 H/T-$NbSe_2$ junction. The STM tip is positioned on the graphene surface near the edge of a T-$NbSe_2$ island, with a tunneling current of 1 nA and a bias voltage of -0.1 V. The tip is then moved along a preset trajectory (indicated by the arrow) to laterally push the T-$NbSe_2$ island toward an H-$NbSe_2$ island with nanoscale priscion, ultimately forming an H/T-$NbSe_2$ junction. The facile translation of $NbSe_2$ islands on graphene is enabled owing to the low mechanical friction between $NbSe_2$ and graphene [24,25]. Simultaneously, STM imaging allows the manipulation process to be monitored in real time, as typically presented in Figs. 2(c) and (d) for the STM images acquired before and after the manipulation. Using this approach, the H-$NbSe_2$ and T-$NbSe_2$ islands are brought into contact at a well-defined but structurally discontinuous interface. Each island retaining its intrinsic edge structures, without atomic bonding across the interface [26,27], as clearly shown in Fig. 2(e).

Figure 2(f) presents site-dependent STS spectra acquired near the type-1 H/T-$NbSe_2$ junction at the positions marked in Fig. 2(e). As the tip moves from H-$NbSe_2$ across the interface into T-$NbSe_2$, the intrinsic superconducting gap of H-$NbSe_2$, approximately 1.8 meV, is abruptly reduced to about 1.1 meV at the interface, and can persist inside the T-$NbSe_2$ region over several SOD lengths, indicating a finite superconducting proximity effect across the nominally insulating phase. To elucidate the origin, we perform large-bias STS measurements across the H/T-$NbSe_2$ junction, as shown in Fig. S3 [20]. The spectra reveal a local suppression of the strongly correlated insulating gap of T-$NbSe_2$ near the interface. This suppression can be attributed to enhanced screening by itinerant carriers in the metallic H-$NbSe_2$, which reduces the effective Coulomb interaction in the adjacent T-$NbSe_2$ and hence closes the local gap [28]. Consequently, a metallic channel emerges along the edge of T-$NbSe_2$, providing an effective pathway for the proximity-induced penetration of Cooper pairs. Considering the structurally discontinuous nature of the type-1 H/T-$NbSe_2$ junction, these results indicate that superconductivity is induced in a charge-transfer insulator through a conventional proximity effect under weak interfacial electronic coupling.

The type-2 H/T-$NbSe_2$ junction, characterized by atomically continuous interfaces, can be constructed through a tip-pulse-induced local phase transition, as schematically illustrate in Figs. 3(a) and (b) [29,30]. Figures 3(c) and (d) show representative STM images of an H-$NbSe_2$ island before and after application of a 5 V tip pulse with a duration of 20 ms. The pulse can induce local phase transitions at two corners of the island, converting H-$NbSe_2$ into T-$NbSe_2$, thus forming an atomically continuous H/T-$NbSe_2$ junction (Fig. 3(e)).

Spatially resolved STS spectra acquired across the type-2 H/T-$NbSe_2$ junction reveal a significantly distinct superconducting proximity effect. As shown in Fig. 3(f) and Fig. S4 [20], the superconducting gap measured on T-$NbSe_2$ near the interface is almost identical to that of intrinsic H-$NbSe_2$, indicating a strong superconducting proximity effect. Moreover, a pair of pronounced resonance peaks emerges inside the gap, symmetrically located with respect to $E_F$, and extends over a narrow spatial range of approximately two SODs from the interface. We can rule out the defects, impurities, or structural disorder as the origin of these in-gap states, since the atomic-resolution STM image in Fig. 3(e) shows no corresponding signatures. Notably, these in-gap states are absent in the weak interfacial electronic coupling regime shown in Fig. 2.

Previous studies have demonstrated that each SOD in monolayer T-$NbSe_2$ hosts one unpaired electron and thus carries a local spin. The observed in-gap resonances are therefore reminiscent of Yu-Shiba-Rusinov (YSR) states, which arise from exchange coupling between localized magnetic moments and Bogoliubov quasiparticles in a superconductor [31]. To examine this scenario, we first establish the presence of local magnetic moments in T-$NbSe_2$ through the Kondo effect (Fig. 4(a) and (b)). Figure 4(d) shows site-resolved STS spectra acquired at 4.2 K on T-$NbSe_2$ supported by metallic H-$NbSe_2$ at the locations marked in Fig. 4(c). In the interior of T-$NbSe_2$, the spectra display a remarkable zero-bias Kondo resonance, evidencing local magnetic moments associated with individual SODs, consistent with previous results [17,18,32]. Within approximately two SODs from the T-$NbSe_2$ edge, however, the Kondo resonance is vanished. Instead, the spectra recover the characteristic features of monolayer T-$NbSe_2$, including the UHB and LHB&VB, while retaining a finite DOS between them. This behavior indicates that, near the T-$NbSe_2$ edge, the local-spin character is still retained, whereas the H/T-$NbSe_2$ interlayer coupling remains weak.

Taken together, the observed superconducting proximity effect accompanied by in-gap resonances in the type-2 H/T-$NbSe_2$ junction can be understood as a cooperative consequence of the interfacial metallic state, which is rooted by the reduced effective Coulomb interaction, interfacial charge transfer, and orbital hybridization. This metallic channel provides a finite DOS at $E_F$, further enabling exchange coupling between the local spins in T-$NbSe_2$ and the penetrating Cooper pairs from H-$NbSe_2$, and thereby giving rise to the YSR states. Theoretically, the YSR state energy is given by $\varepsilon = \Delta\frac{1-(JS\pi\rho)^2}{1+(JS\pi\rho)^2}$, where $\Delta$ is the superconducting gap, $S$ is the impurity spin, $\rho$ is the normal-state DOS at $E_F$, and $J$ is the exchange coupling [33]. This expression indicates that the formation of YSR states requires three essential ingredients: a finite DOS at $E_F$, local magnetic moments, and strong exchange coupling to the superconductor. All of these conditions are fulfilled in the type-2 H/T-$NbSe_2$ junction with a strong interfacial electronic coupling regime.

Previous studies have predominantly investigated YSR states induced by magnetic atomic or molecular impurities adsorbed on superconducting surfaces [34-38], and more recently, have extended these studies to vertical van der Waals heterostructures comprising magnetic materials on superconductors [33,39-41]. In contrast, our results provide the first experimental evidence for the emergence of YSR states at H/T-$NbSe_2$ lateral junctions, establishing a distinct platform where superconductivity, interfacial metallicity, and correlation-driven local moments coexist within a spatially confined interfacial region.

In summary, we systematically investigate proximity effect between superconducting monolayer H-$NbSe_2$ and charge-transfer insulating monolayer T-$NbSe_2$. Atomically sharp monolayer H/T-$NbSe_2$ lateral junctions with distinct interfacial coupling strengths are controllably constructed *in situ* through STM tip manipulation. In the weak-coupling regime, T-$NbSe_2$ develops a robust and markedly reduced superconducting gap, arising from the proximity-induced penetration of Cooper pairs from H-$NbSe_2$. In the strong-coupling regime, by contrast, particle-hole-symmetric YSR bound states emerge within the induced superconducting gap, originating from exchange coupling between local spins in T-$NbSe_2$ and the penetrating Cooper pairs. These findings establish H/T-$NbSe_2$ lateral junctions as a tunable platform for exploring interfacial superconductivity and correlation-driven bound states, which may serve as building blocks for realizing topological superconductivity.

## Acknowledgments

This work was supported by the National Key R&D Program of China (Grant Nos. 2022YFA1402502, 2022YFA1402602), National Natural Science Foundation of China (Grant Nos. 92580103, 12274026, 12321004), Open Fund of Key Laboratory of Multiscale Spin Physics (Ministry of Education), Beijing Normal University (Grant No. SPIN2025K01), Young Elite Scientists Sponsorship Program of the Beijing High Innovation Plan (Grant No. 20250836), and Beijing Institute of Technology Kunpeng&Ascend Center of Cultivation.

## Competing interests

The authors declare no competing interest.

## Data Availability

The Source Data underlying the figures of this study are available. All raw data generated during the current study are available from the corresponding authors upon request.

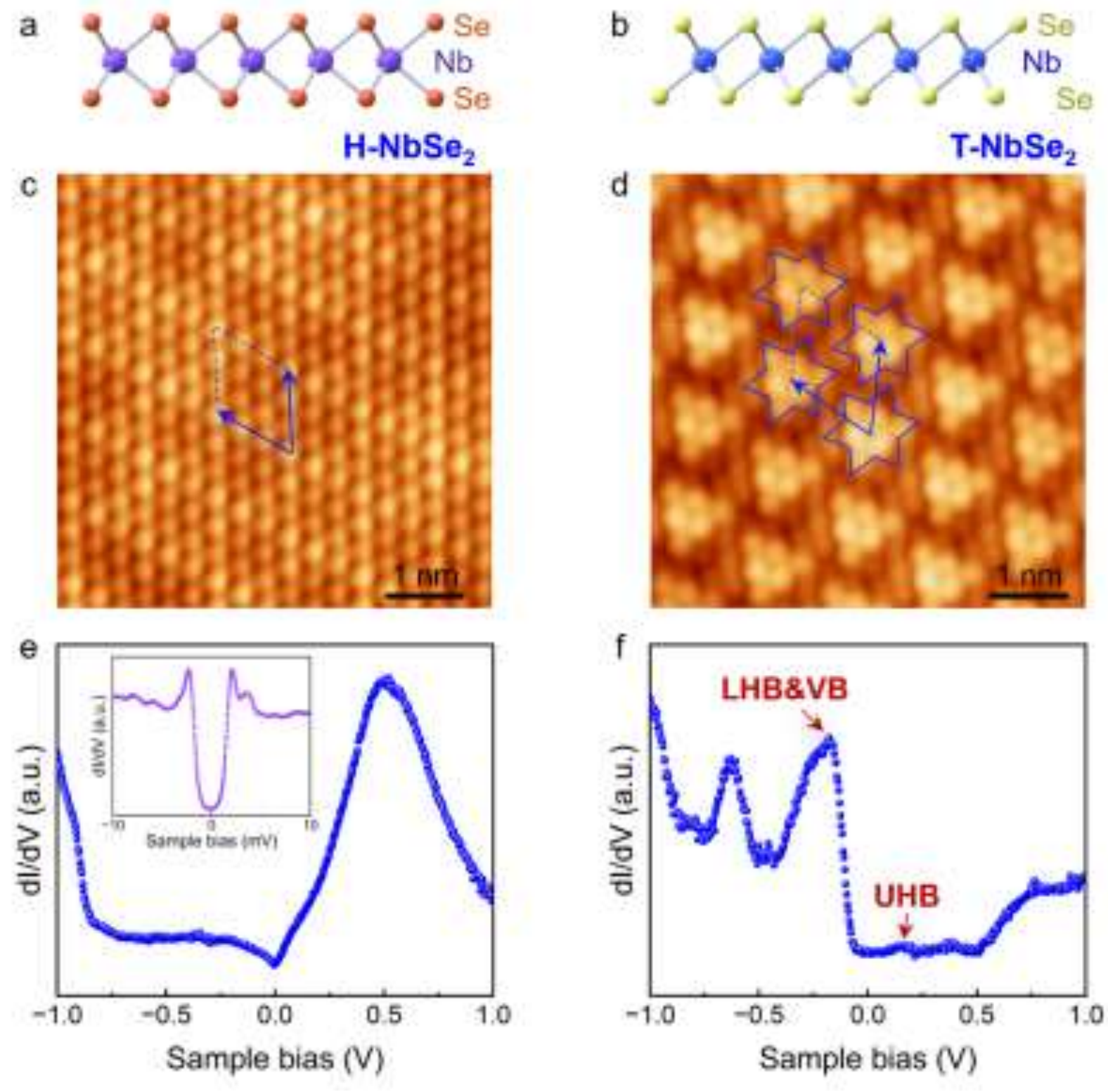

**Figure 1. Structural and electronic properties of monolayer H-NbSe$_2$ and T-NbSe$_2$.** (a,b) Atomic structures of monolayer H-NbSe$_2$ and T-NbSe$_2$, respectively. (c) Atomically resolved STM image of monolayer H-NbSe$_2$ ($V_b$ = −1.0 V, $I_t$ = 300 pA), revealing a 3 × 3 CDW superlattice. (d) Atomically resolved STM image of monolayer T-NbSe$_2$ ($V_b$ = −1.5 V, $I_t$ = 100 pA), showing a $(\sqrt{13} \times \sqrt{13})R13.9°$ CDW superlattice. (e) Representative STS spectrum of monolayer H-NbSe$_2$, showing a CDW gap near the $E_F$. Inset: high-resolution dI/dV spectrum of monolayer H-NbSe$_2$ acquired at 350 mK, exhibiting a superconducting gap symmetric about $E_F$. (f) Representative STS spectrum of monolayer T-NbSe$_2$, with the UHB and LHB&VB orbitals labeled.

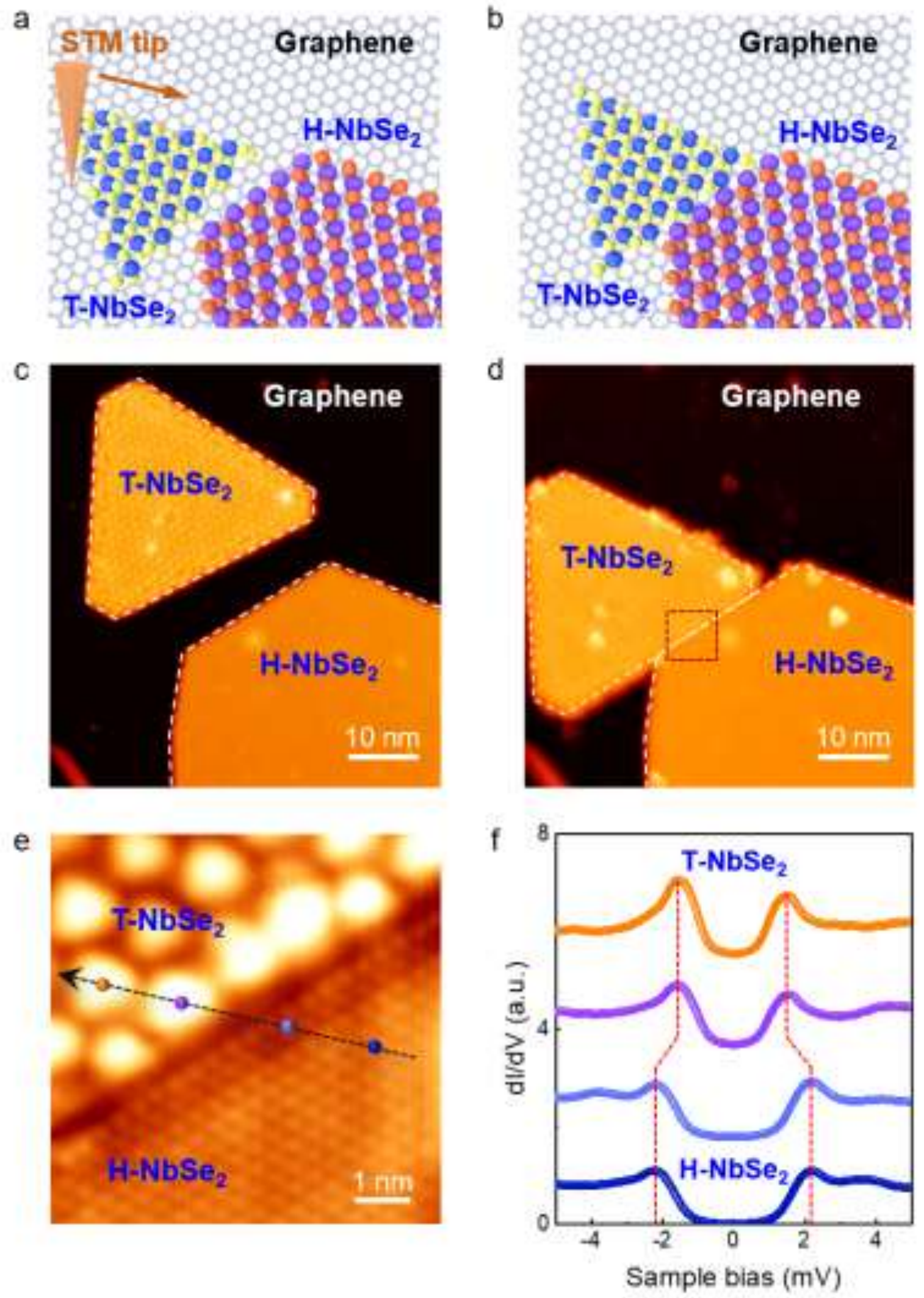

**Figure 2. Superconducting proximity effect in a charge-transfer insulator via a weak interfacial electronic coupling.** (a,b) Schematic illustration of tip-pushed lateral displacement of $NbSe_2$ islands. The STM tip is positioned on graphene near the T-$NbSe_2$ island and then moved along the preset trajectory indicated by the arrow. During this process, the T-$NbSe_2$ island slides on graphene toward the H-$NbSe_2$ island, ultimately forming a H/T-$NbSe_2$ lateral junction (type-1). (c) Large-scale STM image of isolated monolayer T-$NbSe_2$ and H-$NbSe_2$ islands on graphene ($V_b = -1.0$ V, $I_t = 20$ pA). (d) STM image acquired after tip manipulation of the T-$NbSe_2$ island, showing that the two islands are in contact while retaining their original configurations ($V_b = -1.0$ V, $I_t = 20$ pA). (e) High-resolution STM image of the H/T-$NbSe_2$ interface marked in panel d ($V_b = -0.8$ V, $I_t = 100$ pA). (f) Site-dependent STS spectra acquired at the locations marked in panel e. The spectra are vertically offset for clarity.

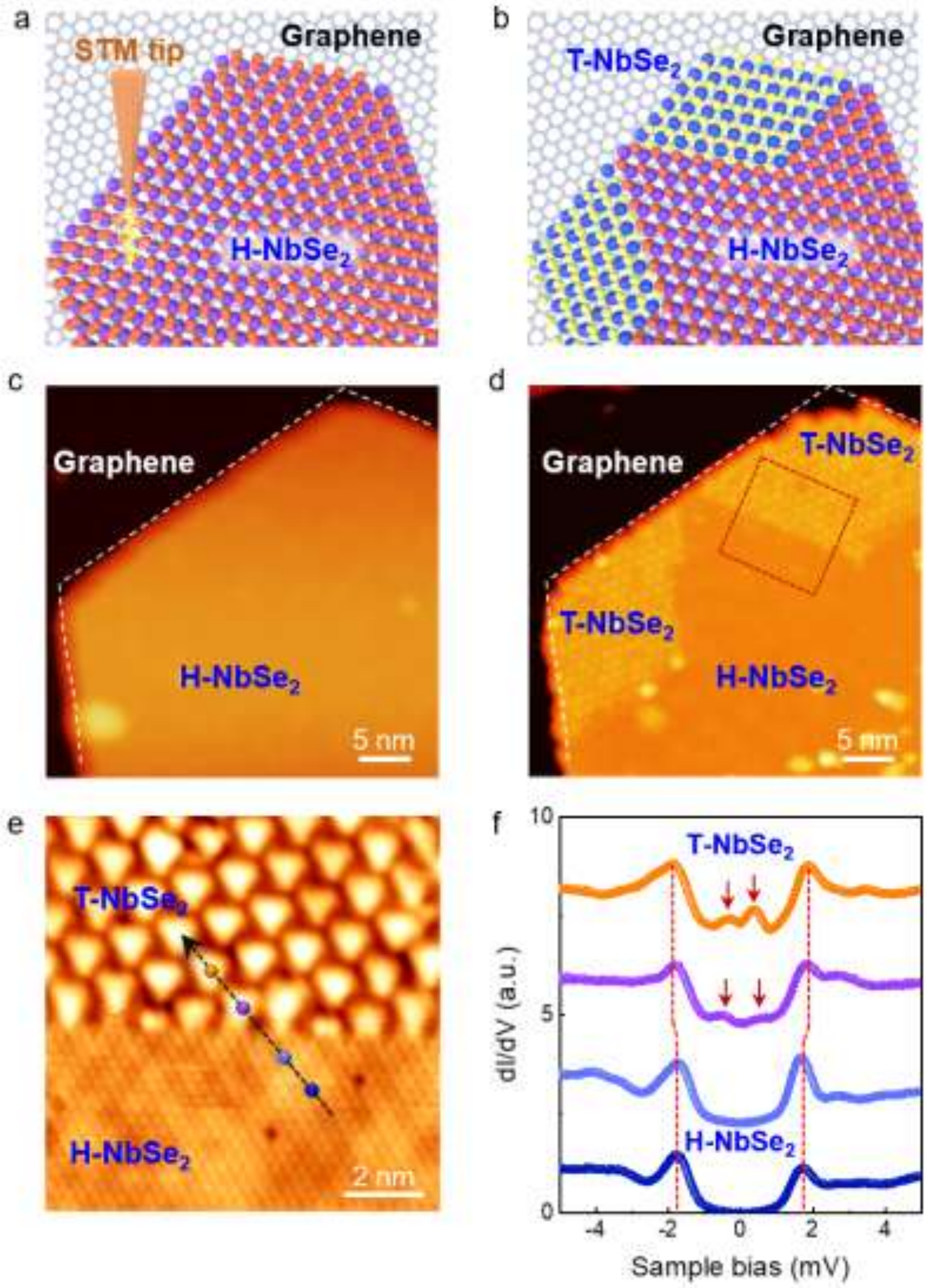

**Figure 3. Superconducting proximity effect in a charge-transfer insulator via a strong interfacial electronic coupling.** (a,b) Schematic illustration of tip-pulse-induced local phase transition process. Upon application of an STM tip pulse, a portion of the H-$NbSe_2$ locally transforms into T-$NbSe_2$, yielding atomically continuous H/T-$NbSe_2$ junctions (type-2). (c) Large-scale STM image of a H-$NbSe_2$ island on graphene ($V_b = -1.5$ V, $I_t = 10$ pA). (d) STM image acquired after applying a 5 V tip pulse to H-$NbSe_2$ for 50 ms ($V_b = -1.5$ V, $I_t = 10$ pA). A portion of H-$NbSe_2$ undergoes a phase transition into T-$NbSe_2$ while the atomic lattice remains intact. (e) High-resolution STM image of the type-2 H/T-$NbSe_2$ junction marked in panel d ($V_b = -0.8$ V, $I_t = 100$ pA). (f) Site-dependent STS spectra acquired at the locations marked in panel e. The spectra are vertically offset for clarity.

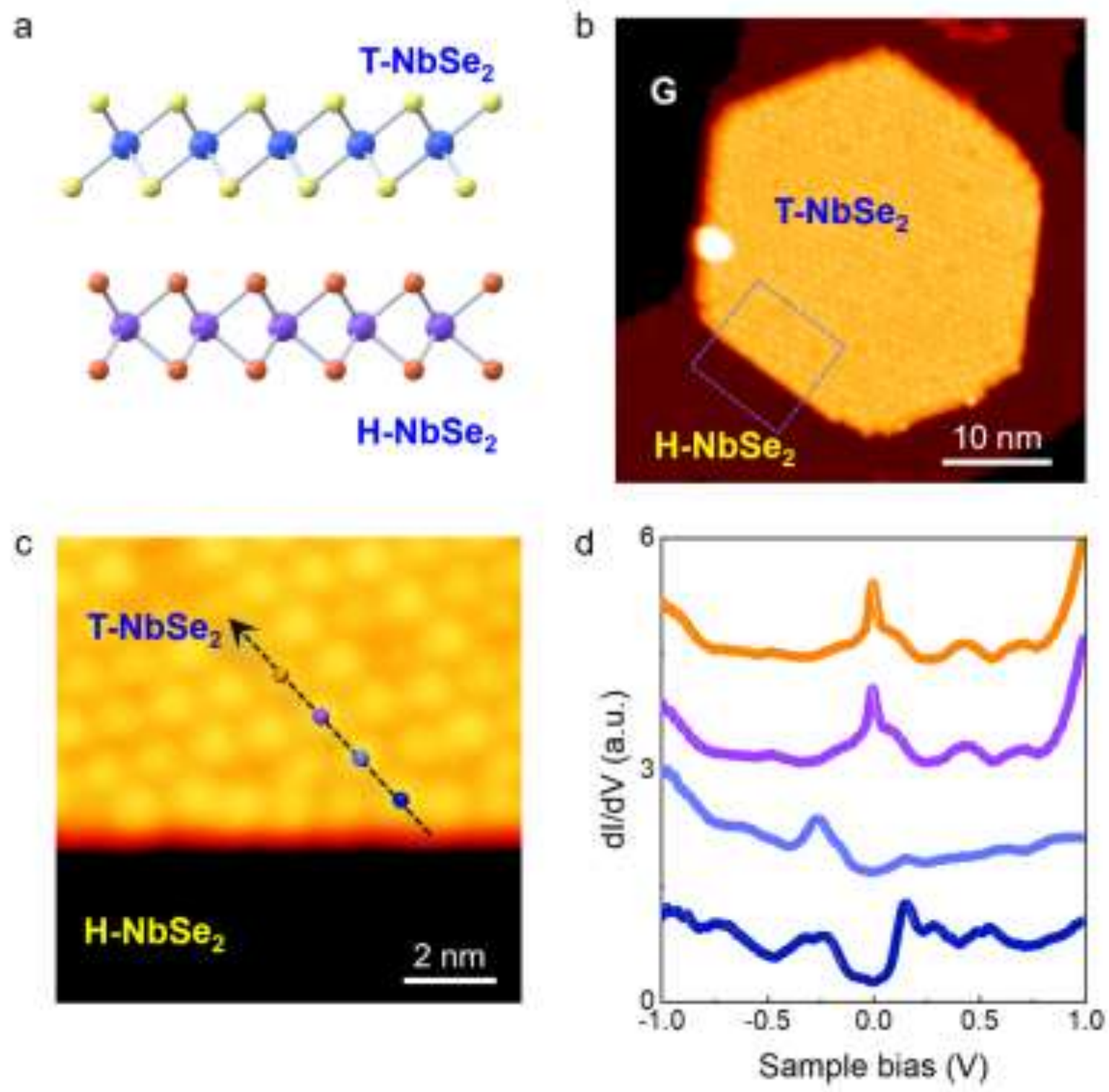

**Figure 4. Local mangetic moments in a charge-transfer insulator.** (a) Atomic structure of a monolayer T-$NbSe_2$ island on top of a monolayer H-$NbSe_2$ island. (b) Large-scale STM image of a T-$NbSe_2$ island supported by a metallic H-$NbSe_2$ island on graphene ($V_b = -1.5$ V, $I_t = 10$ pA). (c) High-resolution STM image of the T-$NbSe_2$ island edge marked in panel b. (d) Site-dependent STS spectra acquired at the locations marked in panel c under the temperature of 4.2 K. The spectra are vertically offset for clarity.